\documentclass[12pt]{article}

\usepackage{epsfig,amsmath,amssymb,latexsym}

\providecommand{\beqa}{\begin{eqnarray}}
\providecommand{\eeqa}{\end{eqnarray}}
\providecommand{\Om}{{\Omega}}
\providecommand{\Ombar}{\overline{\Omega}}
\providecommand{\Ibar}{\overline{I}}

\providecommand{\Wbar}{\overline{W}}

\providecommand{\we}{\wedge}
\providecommand{\tr}{\text{tr}}
\providecommand{\sS}{\sigma_S}
\providecommand{\sT}{\sigma_T}

\def\Orb{{\mathbf{S}^1/\mathbf{Z}_2}}
\def\Z2{{\mathbf{Z}_2}}
\def\mX{{\mathbf{X}}}

\def\cV{{\cal{V}}}
\def\Re{{{\text{Re}}}}

\numberwithin{equation}{section}

\begin{document}

\thispagestyle{empty}
\rightline{LMU-ASC 41/06}
\rightline{HUTP-06/A0020}

\vspace{0.5truecm}

\begin{center}
{\bf \LARGE $\boldsymbol{S}$-$\boldsymbol{T}$rack Stabilization of}\\
\vspace{.4truecm}
{\bf \LARGE Heterotic de Sitter Vacua}\\
\end{center}

\vspace{1.3truecm} \centerline{Gottfried
Curio$^{a,}$\footnote{curio@theorie.physik.uni-muenchen.de} and
Axel Krause$^{b,c,}$\footnote{akrause@fas.harvard.edu}}

\vspace{.7truecm}

\centerline{\em $^a$Arnold-Sommerfeld-Center for Theoretical
Physics}
\centerline{\em Department f\"ur Physik,
Ludwig-Maximilians-Universit\"at M\"unchen}
\centerline{\em Theresienstra\ss e 37, 80333 M\"unchen, Germany}

\vspace{.2truecm}

\centerline{\em $^b$Jefferson Physical Laboratory, Harvard
University, Cambridge, MA 02138, USA}

\vspace{.2truecm}

\centerline{\em $^c$George P.~\& Cynthia W.~Mitchell Institute
for Fundamental Physics}
\centerline{\em Texas A\&M University, College Station, TX 77843, USA}

\vspace{1.0truecm}


\begin{abstract}
\noindent
We present a new mechanism, the $S$-$T$rack, to stabilize the volume modulus $S$ in heterotic M-theory flux compactifications along with the orbifold-size $T$ besides complex structure and vector bundle moduli stabilization. The key dynamical ingredient which makes the volume modulus stabilization possible, is M5-instantons arising from M5-branes wrapping the whole Calabi-Yau slice. These are natural in heterotic M-theory where the warping shrinks the Calabi-Yau volume along $\Orb$. Combined with $H$-flux, open M2-instantons and hidden sector gaugino condensation it leads to a superpotential $W$ which stabilizes $S$ similar like a racetrack but without the need for multi gaugino condensation. Moreover, $W$ contains two competing non-perturbative effects which stabilize $T$. We analyze the potential and superpotentials to show that it leads to heterotic de Sitter vacua with broken supersymmetry through non-vanishing F-terms.
\end{abstract}

\vspace{1.4cm}
\noindent

\newpage
\pagenumbering{arabic}

\section{Introduction}

One of the most interesting regimes of string-theory which allows
to directly address open questions in relevant grand unified
theories or early universe cosmology, is the heterotic string, in
particular its strongly coupled regime, M-theory on
$\mathbf{S}^1/\mathbf{Z}_2$. For non-zero string coupling constant
$g_s$ a new eleventh dimension opens up, the orbifold interval
$\mathbf{S}^1/\mathbf{Z}_2$, whose size $L\sim g_s^{2/3}$
geometrizes the dilaton. The 11-dimensional spacetime, which
incorporates the weakly coupled limit when $L\rightarrow 0$, is
bounded by two 10-dimensional $\mathbf{Z}_2$ fixed planes, the
boundaries, constituting the visible and the hidden sector. Both
generate a non-trivial four-form $G$-flux which leads to a warping
of the background geometry \cite{Witten:1996mz}, \cite{Curio:2000dw}.
When compactified further on a Calabi-Yau threefold $\mathbf{X}$
down to four dimensions to give an effective $N=1$ supergravity,
the warping decreases the size of $\mathbf{X}$ along
$\mathbf{S}^1/\mathbf{Z}_2$. For this compactification of M-theory on a warped seven-manifold $\mathbf{X}\times\mathbf{S}^1/\mathbf{Z}_2$ with $SU(3)$ structure fibered along $\mathbf{S}^1/\mathbf{Z}_2$, $\mX$ becomes a conformally deformed Calabi-Yau whose conformal deformation is determined by the warp-factor.

Starting with the seminal work \cite{Witten:1996mz}, it became clear
that the hidden boundary should be located close to the point where the warp-factor vanishes and the volume of $\mathbf{X}$ would shrink to zero size, classically. It has then been shown in  \cite{Curio:2001qi}, \cite{Becker:2004gw} that the non-perturbative dynamics of the theory indeed stabilizes the  $\mathbf{S}^1/\mathbf{Z}_2$ size and therefore the hidden boundary close to this critical location in vacua with positive energy density and spontaneous breaking of the $N=1$ supersymmetry. The stability of the resulting de Sitter vacua under the inclusion of higher order $R^4$ corrections has been established in \cite{Anguelova:2005jr}. Complementary, stable anti-de Sitter vacua have been investigated in \cite{Buchbinder:2003pi} and were later ``uplifted'' to de Sitter vacua in \cite{Buchbinder:2004im}. The warped background together with these stabilization mechanisms have played an essential role in recent cosmological applications of M-theory ranging from inflation \cite{Becker:2005sg}, \cite{Buchbinder:2004nt} to the creation of heterotic cosmic strings \cite{Becker:2005pv},
\cite{Buchbinder:2005jy} and the potential solution of the strong CP problem via M-theory axions and warping \cite{Svrcek:2006yi}.

While the stabilization of the complex structure moduli and the
K\"ahler moduli is by now quite well understood, little is known
about the stabilization of the $S$ modulus whose real part
measures the size of the average $\mathbf{X}$ volume. Since
non-K\"ahler compactifications cannot exist away from the
boundaries, they are not an option in heterotic M-theory
\cite{progress} and the $S$ modulus stabilization remains a
challenging problem of central phenomenological importance.
It is the goal of this article to show that its stabilization follows
in fact naturally from including the non-perturbative dynamics of
M5-instantons. They arise from Euclidean M5-branes wrapping $\mathbf{X}$. Since the warping of the heterotic M-theory background shrinks the size of $\mX$ along the $\Orb$ interval, the relevance of M5-instantons becomes particularly clear in this theory. Together with hidden sector gaugino condensation and its induced $H$-flux, the M5-instantons provide a superpotential whose $S$ dependent part has a mathematical structure close to a racetrack. Hence it will stabilize $S$ similar to the racetrack. On the other hand, its $T$ dependent part contains two opposing open M2-instanton and gaugino condensate effects which stabilize $T$. We will call this $S$ and $T$ stabilizing mechanism because of its similarity with the ordinary racetrack an $S$-$T$rack. It is noteworthy to stress that it does require multi gaugino condensation. Its key characteristic is the very economical double role played by gaugino condensation which participates both in the $S$ and $T$ stabilization.

\section{Fixing Complex Structure Moduli at High Scale}

Let us start by discussing the stabilization of the complex
structure moduli. They will be fixed in the full 11-dimensional theory before going to the effective 4-dimensional $N=1$ supergravity approximation. The warping of the compactification geometry \cite{Witten:1996mz}, \cite{Curio:2000dw} reduces the volume of the hidden $\mathbf{X}$ slice and thus renders the hidden gauge theory strongly coupled at high energies. Hence we should add a hidden sector gaugino condensate to the full 11-dimensional action. This has the consequence that the 11-dimensional action acquires the following perfect square \cite{Dine:1985rz}, \cite{Horava:1996vs}
\beqa
\int d^{11}x \sqrt{-g}  \Big(G_{lmn11}-\frac{\sqrt{2}}{16\pi}
\left(\frac{\kappa}{4\pi}\right)^{2/3}
\delta(x^{11}-L){\bar \chi}^a \Gamma_{lmn}\chi^a\Big)^2 \; ,
\eeqa
which combines $G$-flux and the gaugino condensate localized on the hidden boundary. Here $\kappa$ is the 11-dimensional gravitational coupling constant. On shell, after setting the variation of the action to zero, this square has to vanish. Hence the hidden sector gaugino
condensate will induce a non-vanishing NS three-form $H$-flux $G_{lmn11} \sim H_{lmn} \delta(x^{11}-L)$ in the hidden sector. The resulting vacuum energy is classically zero and all complex structure moduli are stabilized at a high scale through the alignment of $\Lambda^3\Omega + \bar\Lambda^3\bar\Omega$ with the $H$-flux via the gaugino
condensate \cite{Dine:1985rz}, \cite{Gukov:2003cy}
\beqa
H \sim \langle \tr {\bar \chi}^a \Gamma^{(3)}\chi^a \rangle
= g_{YM,h}^2 (\Lambda^3\Omega + \bar\Lambda^3\bar\Omega) \; .
\eeqa
Notice that the gaugino condensate has to be proportional to the
Calabi-Yau's holomorphic three-form $\Omega$ plus its complex conjugate \cite{Dine:1985rz} which encode the complex structure moduli. Furthermore \cite{Shifman:1987ia}
\beqa
\bar\Lambda^3 = 16\pi^2 M_{UV}^3 e^{-f_h/C_H}
\; , \qquad f_h = S - \beta T
\eeqa
is the scale at which the hidden sector gauge coupling becomes
strong, $f_h$ is the gauge kinetic function of the hidden sector,
$C_H$ the dual Coxeter number of the hidden gauge group $H$ and
$M_{UV}$ its ultraviolet cutoff. The slope parameter which controls the $T$ correction to $f_h$ can be expressed as \cite{Krause:2007gj}
\beqa
\beta = L/L_c \; .
\eeqa
Hence $\beta$ will be of order one in the strongly coupled regime
($L\lesssim L_c$) whereas in the weakly coupled regime ($L\rightarrow
0$) $\beta T$ appears as a small 1-loop correction.

The stabilization of the complex structure moduli happens in the full 11-dimensional theory at a high energy scale. When we investigate below the 4-dimensional effective theory which captures only the massless modes, the complex structure moduli can therefore be regarded as decoupled or more precisely ``integrated out'' \cite{deAlwis:2005tg} and enter the effective theory with fixed values. In fact at quantum level there is an obstruction to setting
$H$ equal to the exponentially small condensate as done above. The
obstruction stems from the observation made in \cite{Rohm:1985jv}
that the $H$-flux has to be quantized. A solution to this puzzle
for general $\mathbf{X}$ comes from the warping of the heterotic M-theory background \cite{Krause:2007gj}\footnote{A proposal based on Calabi-Yau's supporting fractional Chern-Simons invariants appeared in \cite{Gukov:2003cy}.}. Namely it turns out that the $H$-flux contribution to the perfect square picks up an additional factor
\beqa
e^{3f(L)} = |(L_c-L)/L_c|
\eeqa
related to the warp-factor $e^{2f(L)}$. Here $L_c$ denotes the critical length at which the warp-factor vanishes, classically. By now there is lots of evidence, starting with  \cite{Witten:1996mz}, that the hidden boundary position $L$ has to be located close to $L_c$ and that in fact the dynamics of the theory accomplishes this \cite{Curio:2001qi}, \cite{Buchbinder:2003pi}, \cite{Becker:2004gw},    \cite{Buchbinder:2004im}. We thus see that the integer-valued $H$-flux becomes suppressed and continuous due to the warp-factor. On the other hand the warp-factor dependence cancels out of the gaugino condensate, hence allowing for a balancing of both contributions \cite{Krause:2007gj}.

For later use let us write down the superpotentials induced by these effects in the effective 4-dimensional theory. The $H$-flux on the hidden boundary leads to a flux superpotential \cite{Gukov:1999gr}
\beqa
W_H = \frac{e^{3f(L)}}{M_{Pl}} \sqrt{\frac{8}{V_6}}\int_{\mX_0}
H\wedge\Omega \;,
\eeqa
where $V_6 = \int_{\mX_0} d^6y \sqrt{g_{\mX_0}}$ is the volume of the unwarped Calabi-Yau $\mX_0$. The influence of the warped background manifests itself in the additional warp factor pulled out in front
\cite{Krause:2007gj} (see also \cite{Anguelova:2006qf}). Since the complex structure moduli, on which $W_H$ depends, have been integrated out, $W_H$ will enter the 4-dimensional theory as a constant whose size gets reduced through the warp-factor. The gaugino condensate on the other hand generates a superpotential \cite{Dine:1985rz}, \cite{Lukas:1997rb}
\beqa
W_{GC} = g e^{-f_h/C_H} \; , \qquad g = -C_H \mu^3 \; ,
\eeqa
whose scale is given by $\mu \simeq (2M_{GUT}/M_{Pl}) / (32\pi^2)^{1/3} \simeq 3.7 \times 10^{-3}$ \cite{Becker:2004gw}.
For a recent analysis of the interplay of $H$-flux and gaugino condensate see \cite{Manousselis:2005xa}.

\section{Low Energy Dynamics}

Let us next describe the remaining unfixed moduli which enter
the low-energy description after a Kaluza-Klein dimensional
reduction of the 11-dimensional theory. For simplicity we will
choose $\mathbf{X}$ to have Hodge number $h^{(1,1)}=1$ since it is
straightforward to generalize our stabilization mechanism to
general $h^{(1,1)}$. The low energy moduli comprise the complexified volume and orbifold size moduli
\begin{alignat}{3}
S &= \frac{s}{2}+i\sS \; , \qquad\quad \frac{s}{2} = \cV \\
T &= \frac{t}{2}+i\sT \; , \qquad\quad \frac{t}{2} = \cV_{OM} \; .
\end{alignat}
$\cV$ denotes the $\mathbf{X}$ volume averaged over $\Orb$.
With $h^{(1,1)}=1$ there is only one holomorphic 2-cycle on $\mathbf{X}$ and $\cV_{OM}$ denotes the size of a 3-cycle composed out of this 2-cycle $\Sigma_2$ (of average 2-cycle size $\cV^{1/3}$) and the interval $\Orb$. Hence $\cV_{OM}\sim L$ measures the size of $\Orb$. We refer the reader to \cite{Becker:2004gw} for a more  detailed description of these moduli and the physical origin of their  imaginary parts, the axions.

Besides these geometrical moduli, there are further vector bundle
moduli $\Phi_u, \, u=\text{dim} \, H^1(\mX, \text{End}\, V)$. Part
of their parameter space becomes most palpable for vector bundles
$V$ constructed from a spectral cover surface $C$ in $\mX$ when
the latter is elliptically fibered over a base surface $B$. The
surface $C$ is a ramified finite covering of the base $B$ and
consists fiberwise over $b\in B$ of those points of the elliptic
fiber $E_b$ which represent the line bundles in whose sum $V$
decomposes along $E_b$. The moduli space of such a bundle $V$ is
then (partially) built up just by the deformations of $C$ inside
$\mX$. In a dual $F$-theory picture these moduli are just related
to the complex structure deformations of the $F$-theory fourfold.

The dynamics of these moduli is described in the effective
low energy $N=1$ supergravity by the K\"ahler potential and
superpotential. The K\"ahler potential receives contributions from $S$, $T$, the fixed complex structure moduli $Z_\alpha$ and the vector bundle moduli $\Phi_u$
\beqa
K = K_{(S)}+K_{(T)}+K_{(Z)}+K_{(\Phi)},
\eeqa
which are
\beqa
K_{(S)}=-\ln \left( S+\bar S \right) , \quad
K_{(T)}=-\ln\Big(\frac{d}{6}(T+\bar T)^3\Big) , \quad
K_{(Z)}=-\ln\Big(i\int_{\mX_0} \Omega\wedge\bar\Omega\Big)
\eeqa
and $d$ denotes the Calabi-Yau intersection number. Little is
known about $K_{(\Phi)}$ but fortunately we won't need its detailed
structure for the stabilization of the vector bundle moduli and will henceforth suppress it.

Next to the already discussed superpotentials generated by $H$-flux and gaugino condensate, the total superpotential
\beqa
W = W_H+W_{M2}+W_{GC}+W_{M5}
\eeqa
receives two further contributions from open M2-instantons
\cite{Moore:2000fs}, \cite{Lima:2001jc}, \cite{Carlevaro:2005bk}
and -- essential for our mechanism to stabilize the volume modulus -- from M5-instantons. Not to break supersymmetry explicitly, the open
M2-instantons have to wrap the 3-cycle $\Sigma_2\times\Orb$ and therefore stretch from boundary to boundary. For them to give a
non-vanishing contribution we assume that the holomorphic curve
has genus zero $\Sigma_2 = \mathbf{CP}^1$ \cite{Dine:1987bq},
\cite{Witten:1996bn}. The M5 instantons, on the other hand, stem
from Euclidean M5-branes wrapping the whole threefold $\mX$. They
are particularly well motivated in heterotic M-theory where the warped background shrinks the size of $\mX$ along $\Orb$ and thus leads to an enhancement of the semiclassical exponential instanton amplitude. When the sizes of $\Sigma_2$ and $\mX$ are suitably averaged over $\Orb$, the respective superpotentials appear in four dimensions as
\beqa
W_{M2} = h e^{-T},
\qquad W_{M5} = q e^{-S/4} \; .
\eeqa
The prefactors $h$ and $q$ are 1-loop Pfaffians. $h$ is known to
be a holomorphic function of the complex structure and vector
bundle moduli and can be identified with the Pfaffian of the
chiral Dirac operator of the gauge connection $\text{Pf}(D_-)$
\cite{Buchbinder:2003pi}. It is known that this Pfaffian has a
polynomial dependence on the vector bundle moduli
\cite{Buchbinder:2002ic}, \cite{Buchbinder:2002pr}.

From these ingredients we can now infer the effective 4-dimensional potential. We assume that the vector bundle sector preserves supersymmetry, hence $D_{\Phi}W=0$. The standard F-term expression
then gives us the potential
\begin{alignat}{3}
\label{Potential}
\frac{U}{M_{Pl}^4} &= e^K \Big( \sum_{I,J=S,T,\Phi_u} K^{\Ibar J}
D_{\Ibar}\Wbar D_J W - 3|W|^3 \Big) \\
&= \frac{6}{d}\bigg(
\frac{s}{t^3}\Big|(W_{GC}+W_{M5})_S-\frac{W}{s}\Big|^2
+\frac{1}{3st}\Big|(W_{M2}+W_{GC})_T-\frac{3W}{t}\Big|^2
-\frac{3}{st^3}|W|^2\bigg) \notag \; ,
\end{alignat}
where $i\int_{\mX_0} \Om\we\Ombar = 1$ equals unity in our conventions. We will now come to the moduli stabilization mechanisms.

\section{$\boldsymbol{S}$-$\boldsymbol{T}$rack Moduli Stabilization}

Now, that all dynamical ingredients are given, our aim will be to
demonstrate that the remaining moduli, and in particular the volume
modulus, can be stabilized in de Sitter vacua. We will analyze
the stabilization at the level of the full potential. Notice that
one could also start by solving the supersymmetry conditions, $D_I
W=0$, thus obtaining first AdS vacua which could then later be lifted in a controllable way by D-terms \cite{Burgess:2003ic}. This has been done in heterotic M-theory in \cite{Buchbinder:2004im}, see also \cite{Braun:2006th}. However, such de Sitter vacua are very special. Only by investigating the full potential can one make sure to find all existing de Sitter vacua.

Let us start with the vector bundle moduli. They are fixed through
the dependence of the $W_{M2}$ prefactor $h = \text{Pf}(D_-)$, given by the Pfaffian of the chiral Dirac operator of the gauge connection, on the $\Phi_u$. For instance, in a special case $h$ was explicitly computed as a polynomial in the $\Phi_u$ \cite{Buchbinder:2002ic}, \cite{Buchbinder:2002pr}. The concrete polynomial dependence was evaluated for the instanton given by the rational curve in terms of the base ${\bf P^1}$ of the Hirzebruch surface (which itself constituted the base $B$ of the elliptic fibration of the Calabi-Yau space ${\bf X}$; note that this example had $h^{1,1}=3$). In principle one has to sum up appropriately the contributions from {\em all} rational curves. The polynomial dependence of $W_{M2}$ on the $\Phi_u$ has been argued to fix their values in acceptable regimes in \cite{Buchbinder:2003pi}. We refer the reader to this work for a detailed discussion and concentrate now on the stabilization of the geometrical moduli.

\subsection{K\"ahler Modulus}

Let us first address the dynamical stabilization of the K\"ahler
modulus $T$. An inspection of the total superpotential reveals two
$T$ dependent exponential terms, marked by arrows
\beqa
W = W_H + h e^{-T} + g e^{-S/C_H}e^{\beta T/C_H}
+ q e^{-S/4} \; .
\thinlines
\put(-153,-13){\line(1,0){81}}
\put(-153,-13){\vector(0,1){8}}
\put(-72,-13){\vector(0,1){8}}
\eeqa
They arise from the open M2-instanton and the gaugino condensate and decrease resp.~increase with $t$. Hence one expects a non-trivial minimum at which both effects balance each other. This balancing mechanism for the stabilization of $t$ had been proposed and confirmed in \cite{Becker:2004gw} at the level of the full effective potential. From the sine and cosine dependence of $W$ on $\sigma_T$, it is also clear that generically the axion $\sigma_T$ is stabilized. This had been verified in \cite{Becker:2004gw} and thus the full complex $T$ is stabilized by the competition between open M2-instantons and the $T$ dependent part of gaugino condensation.

To good approximation it had moreover been shown in \cite{Becker:2004gw} that this balancing amounts to setting the partial $T$ derivative to zero
\beqa
W_{M2} \approx W_{GC}
\qquad \Leftrightarrow \qquad
\partial_T W = (W_{M2}+W_{GC})_T \approx 0 \; ,
\label{TCrit}
\eeqa
which can be understood as follows. The potential can be rewritten as
\beqa
\frac{U d}{6 M_{Pl}^4}
= \frac{s}{t^3}
\Big( |D_S W|^2
+\frac{t^2}{3s^2}|\partial_T W|^2
-2\frac{t}{s^2}\Re(W \overline{\partial_T W})
\Big) \; ,
\eeqa
where the $-3|W|^2$ term dropped out due to the no-scale structure of the K\"ahler-potential for $T$. Notice that for large $s,t\gg 1$, which we have to assume for the validity of the supergravity description, the first two terms dominate and the third, potentially negative term, is subleading. This implies that we have a positive potential, thus breaking supersymmetry. Furthermore, the minimization of the potential is to good approximation given by the minimization of the leading two terms. Their minimization implies  setting $\partial_T W\approx 0$ and moreover $D_S W\approx 0$. The latter condition will be discussed in the next section.

Employing just the $\partial_T W\approx 0$ condition, for now,
we remain at leading order in $s,t$ with a manifestly positive expression
\beqa
\label{Potential2}
\frac{U}{M_{Pl}^4} \approx \frac{6s}{t^3 d}
\Big|(W_{GC}+W_{M5})_S-\frac{W}{s}\Big|^2
\eeqa
at the critical point where $T$ gets stabilized. This positivity is the reason why we obtain {\em de Sitter} rather than anti de Sitter vacua after stabilizing $S$ in the next step. The balancing condition, $\partial_T W\approx 0$, implies that
\beqa
D_T W \approx K_T W = -\frac{3}{t} W \ne 0 \; ,
\eeqa
confirming that supersymmetry gets broken spontaneously through non-vanishing F-terms.

\subsection{Volume Modulus}

Having fixed the K\"ahler $T$ modulus, let us now explain the
mechanism to stabilize the complexified volume modulus $S$. In
fact, all contributions to the superpotential which haven't been
used so far conspire in just the right way to give us a
mathematical structure similar to a generalized racetrack \cite{Escoda:2003fa}. The arrows indicate these terms
\beqa
W = W_H + W_{M2} + g e^{-S/C_H} e^{\beta T/C_H}
+ qe^{-S/4} \; .
\thinlines
\put(-190,-13){\line(1,0){173}}
\put(-190,-13){\vector(0,1){8}}
\put(-107,-13){\vector(0,1){8}}
\put(-17,-13){\vector(0,1){8}}
\eeqa
which come from the $H$-flux, the $S$ dependent part of the gaugino condensate and the M5-instantons. We would like to stress the economic double role played by the gaugino condensate and the fact that this structure does not arise from multi gaugino condensation. Due to its different physical origin and the incorporation of $T$, we call this structure the $S$-$T$rack. Because of its mathematical similarity with the generalized racetrack \cite{Escoda:2003fa}, the complex volume modulus $S$ becomes stabilized as in those models which we verify explicitly in numerical examples later on.

To develop a better understanding of how the $S$-$T$rack stabilization works for $S$ and to see the differences to the standard racetrack scheme, let us start with the latter. The standard racetrack \cite{Krasnikov:1987jj} has its physical origin in a multi gaugino condensate based on a product of gauge groups. This generates a
superpotential of type $W = g_1 e^{-S/C_{H_1}} + g_2
e^{-S/C_{H_2}}$ and has been suggested for a stabilization of the
heterotic string dilaton at weak coupling. The $S$-$T$rack analogy to this would be $W = g e^{-S/C_H} e^{\beta T/C_H} + q e^{-S/4}$ when we set $W_H$ to zero. In fact a similar structure arises in the weakly coupled heterotic string from a gaugino condensate combined with
worldsheet instantons and stabilizes K\"ahler moduli there
\cite{Curio:2005ew}. Stabilization of the $S$ modulus would then follow, as in the standard racetrack, from the stationarity of the superpotential \cite{Dine:1999dx}
\beqa
\partial_S W \approx 0 \quad \Leftrightarrow \quad W_{GC} \approx -W_{M5} \; ,
\label{SCrit}
\eeqa
which demands the balancing of $W_{GC}$ with $W_{M5}$. Here $W_{GC}$ and $W_{M5}$ must have opposite signs at the stationary point. The stationary point itself is given by
\beqa
S = \frac{4C_H}{4-C_H}\left(\ln\Big(\frac{4\mu^3}{q}\Big)
+\frac{\beta}{C_H}T \right) \; .
\label{StatPoint}
\eeqa
In contrast to the racetrack, the $S$-$T$rack shows, next to the logarithmic term, also a linear term in $T$ which arises from the
$T$-dependence of $W_{GC}$. Hence the critical value for $S$ depends also on the critical value for $T$. While the goal of the standard racetrack was to stabilize the dilaton at a small value and relied on the presence of only the logarithmic term, the goal of the $S$-$T$rack is to stabilize both volumes, $S$ and $T$, at sufficiently large values where the supergravity description makes sense. Because the $T$ stabilization leads to such large $T$ values \cite{Becker:2004gw}, we cannot neglect the linear term and naturally also $S$ will become large.

If we now add the $H$-flux to build the full $S$-$T$rack
the simple stationarity requirement, $\partial_S W=0$, would fail to
take the flux contribution into account. The natural condition is
rather to demand the vanishing of the K\"ahler covariant derivative which incorporates $W_H$, as we have demonstrated in the previous section. This implies
\beqa
D_S W \approx 0 \quad \Leftrightarrow \quad \frac{W_H}{s}\approx
-\bigg(\frac{W_{GC}}{C_H}+\frac{W_{M5}}{4}\bigg) \; .
\eeqa
To obtain the rhs we have used $s\gg C_H,4$ which is needed to trust the supergravity description and the $T$ balancing condition $W_{M2}\approx W_{GC}$.

The vanishing of $D_SW$ rather than $\partial_SW$ implies two immediate differences to ordinary racetrack stabilization. First, it amounts to a balancing of $H$-flux with the combined effects of gaugino condensation and M5-instantons and no longer a balancing of just the two latter effects. Hence their signs need no longer be strictly opposite as in (\ref{SCrit}). Second, as we will demonstrate explicitly in an example, the case $C_H=4$ which must be excluded in racetrack stabilization because it would send $S\rightarrow \infty$ is now admitted, too. Since $U \propto |D_SW|^2$, the condition $D_S W \approx 0$ will furthermore lead to small positive vacuum energy densities $U \ll M_{Pl}^4$. Still, the $S$-$T$rack mechanism does not solve the cosmological constant problem in view of the approximate nature of the conditions (see \cite{Krause:2000gp} how a warp-factor might improve on this in a similar IIB setup). It will, however, provide us naturally with a positive vacuum energy density $U$ that is significantly smaller than the reduced Planck-scale. This is indeed a necessary consistency requirement which shows that the analysis can indeed self-consistently be carried out within the low-energy effective theory.

\section{Fixing Axions}

Let us now explain why also both axions $\sigma_S$ and $\sigma_T$ are expected to be fixed. For this we take $s$ and $t$ to be fixed as explained before. Our potential consists of three squares of absolute values
\beqa
U \sim |z_1(\sS,\sT)|^2 + |z_2(\sS,\sT)|^2 + |z_3(\sS,\sT)|^2 \; .
\eeqa
Let us first concentrate just on one square and assume that $U \sim
|z(\sS,\sT)|^2$. It is then clear that the phase $\varphi(\sS,\sT)$ of $z = re^{i\varphi}$ gives rise to an unstabilized axion combination as it drops out of the potential and builds a flat direction.
Now with three simultaneous squares, as in our case, the combination of axions $\varphi_n(\sS,\sT)$ which corresponds to the phase of $z_n=r_n e^{i\varphi_n}$ will still remain a flat direction for $|z_n|^2$. In general these three axion combinations $\varphi_n(\sS,\sT), \; n=1,2,3$ will, however, describe different curves in the 2-dimensional $\sS,\sT$ plane which will intersect at isolated points. These points represent minima of the potential and the flatness which had existed in the $\varphi_n(\sS,\sT)$ directions for single squares gets lifted. Indeed, from our explicit expression for the potential (\ref{Potential}), we see that $z_1\ne z_2\ne z_3\ne z_1$ and consequently their phases will generically be different. Hence, we do not expect flat axion directions to remain. This will be confirmed by our actual numerical analysis.

\section{Stabilized de Sitter Vacua: Numerical Examples}

\begin{figure}[t]
\begin{center}
\includegraphics[scale=0.68]{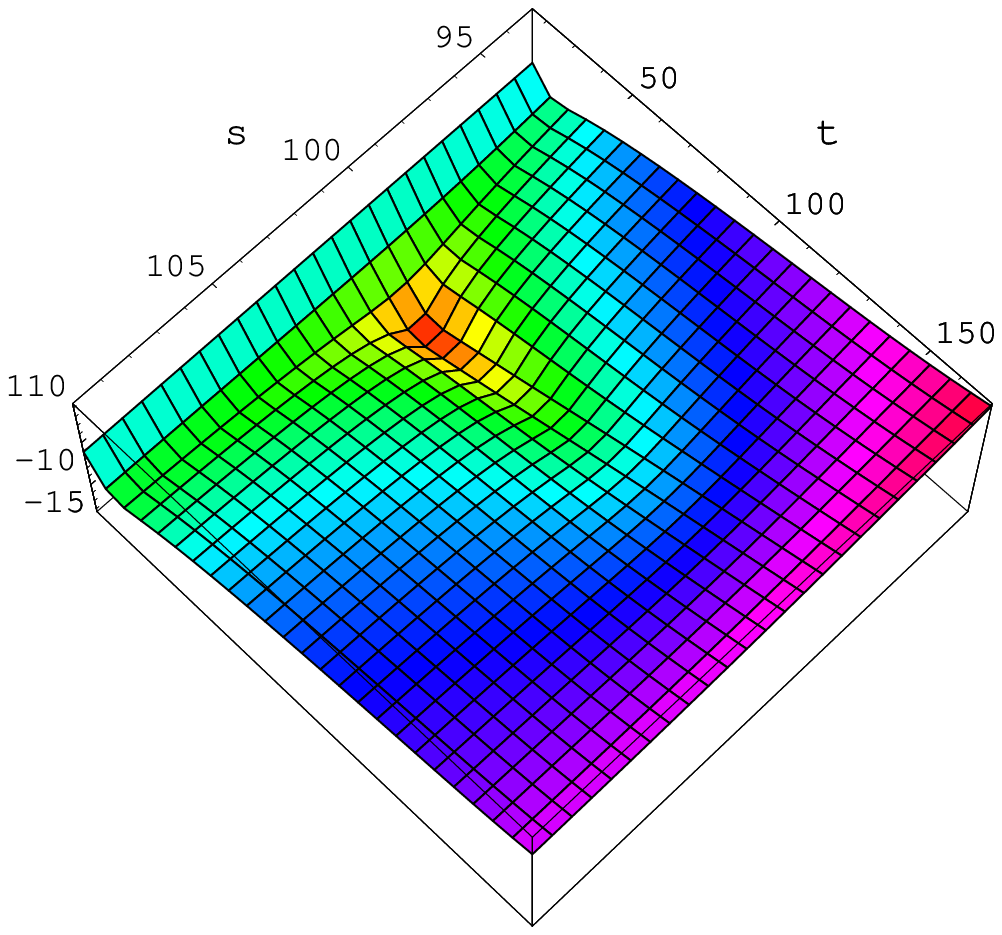}
\includegraphics[scale=0.60]{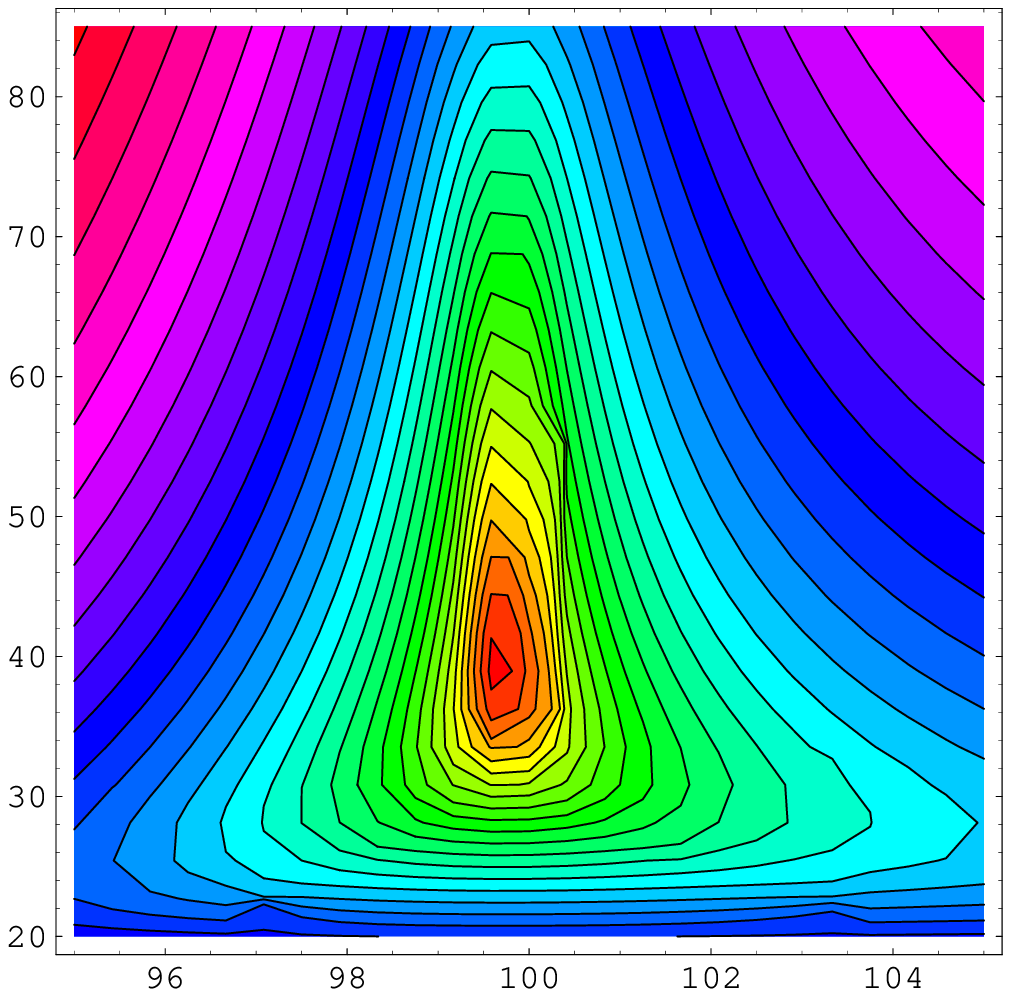}
\caption{\label{Fig1} Plot and contour plot of the logarithm
$\ln(U d/6 M_{Pl}^4)$ of the $S$-$T$rack potential. $s$ is plotted
along the x-axis, $t$ is plotted along the y-axis and we have chosen
the slice $\sS=\sT=0$. Both $t$ and $s$ get clearly stabilized at
values large enough to trust the supergravity description.}
\end{center}
\end{figure}

Having explained the $S$-$T$rack mechanism through which the moduli become dynamically stabilized, let us now present three concrete numerical examples. We solve numerically for stationary points of the full potential by setting
\begin{alignat}{3}
\frac{\partial U}{\partial S} &= \overline{D_SW}
\big(D_SW+s(D_SW)_S\big)
- \frac{t^2}{3s^2} \overline{D_TW} \big(D_TW-s(D_TW)_S\big)
-\frac{2}{s} \Wbar D_SW \notag \\
\frac{\partial U}{\partial T} &=
\overline{D_TW} \big(D_TW-t(D_TW)_T\big)
+ \frac{9s^2}{t^2} \overline{D_SW} \big(D_SW-\frac{t}{3}(D_SW)_T\big)
+ \frac{6}{t}\Wbar D_TW \notag
\end{alignat}
to zero. We will first illustrate how both $t$ and $s$ get stabilized. For this adopt reasonable values
\beqa
\label{values}
\beta = 0.8\;, \qquad W_H = 10^{-4}\;, \qquad
h = 10^{-1}\;, \qquad q = -1
\eeqa
and a hidden gauge group $SO(10)$ which has dual Coxeter number $C_H=8$. Picking the slice $\sS=\sT=0$, we depict in fig.\ref{Fig1} the resulting $S$-$T$rack potential over the $s$-$t$ plane.
A minimum, at values for $s$ and $t$ much larger than one, embedded in a characteristic long straight valley is clearly visible.

Let us next check for stationary points which correspond to minima
with respect to all four $s,t,\sS,\sT$. To find such points
we search the four-plane $(s,t,\sS,\sT)$ for stationary points at which the Hessian of the potential
\beqa
H =
\left(
\begin{array}{cccc}
U_{ss} & U_{st} & U_{s\sS} & U_{s\sT} \\
U_{ts} & U_{tt} & U_{t\sS} & U_{t\sT} \\
U_{\sS s} & U_{\sS t} & U_{\sS\sS} & U_{\sS\sT} \\
U_{\sT s} & U_{\sT t} & U_{\sT\sS} & U_{\sT\sT}
\end{array}
\right)
\eeqa
has all its eigenvalues positive. This search is carried out numerically. Our first example describes a hidden $SU(5)$ with $C_H = 5$ and the same parameters as in (\ref{values}).
The stationary point at
\beqa
s = 164.8\; , \qquad t = 62.7 \; , \qquad \sS = 0
\; , \qquad \sT = 1.9
\eeqa
has all eigenvalues ($\times 10^{14}$) of the Hessian positive:
$2360330, \; 89254.8, \; 8051.6, \; 5.4$. It therefore corresponds to a stabilized minimum. The vacuum energy density at this point is positive.

Our second example describes a hidden $SU(4)$ gauge group with
$C_H=4$ and we switch $\beta$ from 0.8 to 0.9 which brings $L$ closer to $L_c$. For the parameters (\ref{values})
we find now a minimum at
\beqa
s = 164 \; , \qquad t = 70.4 \; , \qquad \sS = 0
\; , \qquad \sT = 1.7
\eeqa
with all eigenvalues ($\times 10^{13}$) of the Hessian positive:
$268701, \; 13968, \; 911.9, \; 3.6$. This shows that the case with hidden gauge group $SU(4)$ which had to be excluded in the standard racetrack shows a regular $S$-$T$rack behavior as argued before. All vacua presented here exhibit non-vanishing $D_T W\ne 0$ and thus break supersymmetry spontaneously.

\bigskip
\noindent {\large \bf Acknowledgements}\\[2ex]
We would like to thank Lubos Motl for interesting related discussions. The work of A.K.~has been supported by the National Science Foundation under grant PHY-0354401 and the University of Texas A\&M.

\bibliographystyle{plain}

\end{document}